# Dispersionless propagation of ultra-short spin-wave pulses in ultrathin yttrium iron garnet waveguides

B. Divinskiy[1*], H. Merbouche[2], K. O. Nikolaev[1], S. Michaelis de Vasoncellos[3], R. Bratschitsch[3], D. Gouéré[2], R. Lebrun[2], V. Cros[2], J. Ben Youssef[4], P. Bortolotti[2], A. Anane[2], S. O. Demokritov[1], and V. E. Demidov[1]

[1]*Institute for Applied Physics, University of Muenster, 48149 Muenster, Germany*

[2]*Unité Mixte de Physique, CNRS, Thales, Université Paris-Saclay, 91767, Palaiseau, France*

[3]*Institute of Physics and Center for Nanotechnology, University of Muenster, 48149 Muenster, Germany*

[4]*LabSTICC, UMR 6285 CNRS, Université de Bretagne Occidentale, 29238 Brest, France*

We study experimentally the propagation of nanosecond spin-wave pulses in microscopic waveguides made of nanometer-thick yttrium iron garnet films. For these studies, we use micro-focus Brillouin light scattering spectroscopy, which provides the possibility to observe propagation of the pulses with high spatial and temporal resolution. We show that, for most spin-wave frequencies, dispersion leads to broadening of the pulse by several times at propagation distances of 10 μm. However, for certain frequency interval, the dispersion broadening is suppressed almost completely resulting in a dispersionless pulse propagation. We show that the formation of the dispersion-free region is caused by the competing effects of the dipolar and the exchange interaction, which can be controlled by the variation of the waveguide geometry. These conclusions are supported by micromagnetic simulations and analytical calculations. Our findings provide a simple solution for the implementation of high-speed magnonic systems that require undisturbed propagation of short information-carrying spin-wave pulses.

*Corresponding author, e-mail: b_divi01@uni-muenster.de

# I. INTRODUCTION

Recent developments in fabrication of nanometer-thick films of low-loss magnetic insulator - yttrium iron garnet (YIG) [1-3] significantly advanced the emerging field of magnonics [4-7], which aims to utilize spin waves in nano-scale magnetic-film structures for transmission and processing of information [8-11]. Due to their small thickness, these films enable fabrication of nano-patterned spin-wave guiding elements with sub-micrometer lateral dimensions [12-15]. Simultaneously, due to the very small magnetic losses, they allow one to achieve large propagation distances of spin waves [16-23] significantly exceeding those in conductive ferromagnetic nano-structures [24]. Additionally, ultrathin YIG films can be efficiently driven by spin-orbit torques, which enables efficient electrical control of propagation characteristics of spin waves [25], as well as excitation of magnetic auto-oscillations [26,27] and spin waves [28] by pure spin currents.

Although the propagation of spin waves in magnetic waveguides based on ultrathin YIG has become a topic of intense research in the recent years [12-15,16-23], the majority of these studies focuses on the continuous-wave propagation regime or addresses propagation of relatively long spin-wave pulses with a duration of several tens of nanoseconds. These works provide important fundamental information about the propagation characteristics. However, since magnetic materials are strongly dispersive media for spin waves [29], the utilization of YIG-based nano-structures for competitive state-of-the-art technical applications requires knowledge about the propagation of short spin-wave pulses with a duration down to a few nanoseconds. Indeed, the information capacity of transmission lines and the operation speed of wave-based computation components are generally limited by dispersion effects resulting in a distortion and temporal overlap of short pulses carrying information, which causes the loss of the latter (see, e.g., Ref. [30]). Therefore, a deep understanding of these phenomena, as well as finding approaches, which allow one to overcome them, is a necessary step in the development of high-speed magnonic nano-circuits.

The present lack of experimental information about the propagation of nanosecond spin-wave pulses in YIG waveguides is largely explained by the absence of a simple experimental technique for their generation. Such pulses can be excited, for example, by using ultra-fast laser systems [31-36], broad-spectrum pulses of magnetic field [37-40], or spin-torque nano-oscillators [41]. The main limitation of all these techniques is the lack of controllability of the carrier frequency of the excited spin waves, which did not allow the systematic analysis of the dispersion effects on the propagation of spin waves with different frequencies and wavelengths.

Here, we study the propagation of spin-wave pulses, which are generated by modulating a continuous-wave high-frequency signal by using ultra-fast commercially available solid-state switches. This provides the opportunity to generate microwave pulses with a well-defined carrier frequency and a duration down to 2.5 ns, limited by the characteristic rise/fall time of the devices. This approach allows us to systematically study the propagation of ultra-short spin-wave pulses in a microscopic 50 nm-thick YIG waveguide. We show that such pulses generally exhibit a strong dispersion broadening up to a factor of three at the propagation distances of 10 μm. More importantly, we find that, within a certain frequency interval, the pulses can propagate without noticeable change of their shape, so that their temporal width remains almost constant. We associate these behaviors with the minimization of the group-velocity dispersion, which is caused by the competing effects of the dipolar and the exchange interaction on the dispersion characteristics of spin waves. This conclusion is supported by analytical calculations, as well as by micromagnetic simulations, showing good agreement with the experimental data. These calculations additionally demonstrate that the spin-wave frequency/wavelength corresponding to the non-dispersive propagation can be controlled by varying the thickness of the YIG film. These findings pave the way for implementation of high-data-rate magnonic circuits that do not suffer from the detrimental dispersion effects.

## II. EXPERIMENT

Figure 1(a) shows the schematics of the experiment. The test device is based on a 50 nm-thick YIG film grown by liquid phase epitaxy on a gadolinium gallium substrate [42]. The values of the saturation magnetization of the film $4\pi M = 1.75$ kG and the Gilbert damping constant $\alpha = 2 \times 10^{-4}$ are measured by using the standard ferromagnetic resonance (FMR) technique. The 1-µm wide YIG waveguide is defined by e-beam lithography using 300 nm-thick PMMA A4 resist, and Ar ion milling through an Au(10nm)/Al(60nm)/Ti(45nm) hard mask fabricated by lift-off. The sputtered gold layer acts as an oxygen barrier, preserving the YIG stoichiometry from the evaporated Al and Ti layers. After the etching, the remaining Au and Al are removed using selective chemical etching (MF-319 developer for Al and KI/$I_2$ for Au). The waveguide is magnetized to saturation by a static magnetic field $H_0$ applied in its plane perpendicular to the long axis. The excitation of spin waves is performed using a 400 nm-wide and 80 nm-thick Au microstrip inductive antenna [43] also defined by e-beam lithography using PMMA A4 resist. To excite ultra-short spin-wave pulses, we pulse-modulate the continuous-wave microwave signal at frequency $f$ by using a high-speed solid-state switch characterized by a rise/fall time of about 1.5 ns. The switch is controlled by voltage pulses with a fixed period of 300 ns produced by a reference pulse generator. Due to the finite reaction time of the switch, the shortest microwave pulses, which can be generated in this way, have the temporal width of about 2.5 ns.

We study the propagation of the spin-wave pulses in the YIG waveguide with spatial and temporal resolution by using micro-focus Brillouin light scattering (µ-BLS) spectroscopy [24]. We focus the probing light with the wavelength of 473 nm and the power of 1 mW produced by a single-frequency laser into a diffraction-limited spot on the surface of the YIG film [Fig. 1(a)]. Due to the interaction with the magnetization in YIG, the probing light is modulated with the frequency equal to that of spin waves. This modulation is analyzed by a six-pass Fabry-Perot interferometer. The resulting signal – the BLS intensity – is proportional to the intensity of the spin wave at the position of the probing spot. By rastering the probing laser spot over the sample surface, we record spatial maps of the spin-wave intensity with the sub-micrometer spatial resolution. By synchronizing the light detector with the reference pulse generator and using a stroboscopic detection technique, we perform time-resolved measurements, which allow us to record temporal dependences of the spin-wave intensity at every spatial location with a temporal step size of 0.2 ns.

### III. RESULTS AND DISCUSSION

Figures 1(b) and 1(c) show snapshots of the BLS intensity recorded at three different time delays relative to the start of the driving microwave pulse, illustrating the propagation of spin-wave pulses with two different carrier frequencies. To better visualize the propagation of the spin-wave pulses, each BLS map shown in Figs. 1(b) and 1(c) is normalized to the maximum value. As seen from the data of Fig. 1(b) obtained at $f = 4.8$ GHz, the spin-wave pulse with this carrier frequency exhibits typical behaviors expected for the propagation of a short pulse in a strongly dispersive medium. During propagation, the width of the pulse increases by about a factor of three, which is known to be caused by the difference in the velocities of its spectral components. In strong contrast, the pulse with the carrier frequency $f = 4.5$ GHz [see Fig. 1(c)] experiences significantly less pronounced dispersion-induced broadening. At the same propagation distance, its width changes by about 30% only. This indicates that the dispersion effects in the YIG waveguide are strongly dependent on the carrier frequency of spin-wave pulses and can be controlled by the proper choice of the latter.

To characterize these effects in detail, we fit temporal profiles of the spin-wave pulse recorded at different distances from the antenna [Fig. 2(a)] by a Gaussian function, and determine the spatial dependences of the temporal width of the pulse for different carrier frequencies [Fig. 2(b)]. As seen from these data, the spin-wave pulse at $f = 4.5$ GHz (point-up triangles) experiences a moderate broadening from about 2.5 ns at $x = 0$ to about 3.3 ns at $x = 10$ µm, which corresponds to an increase of about 30%. With the increase of the carrier frequency to $f = 4.8$ GHz [circles in Fig. 2(b)], dispersion effects become much stronger: the temporal width of the spin-wave pulse increases threefold upon propagation over a distance of 10 µm. However, further increase of the frequency to $f = 5.2$ GHz [point-down triangles in Fig. 2(b)] leads again to a strong weakening of the dispersion-induced broadening. Remarkably, the temporal width of the pulse with the carrier frequency $f = 5.2$ GHz remains practically unaltered over the entire measured distance, indicating that a nearly non-dispersive propagation regime is reached.

Figure 2(c) summarizes the results obtained for different carrier frequencies in the range $f = 4.5$-5.2 GHz, where spin waves can be efficiently excited by the used inductive antenna [43]. As can be seen from these data, the dispersion broadening factor, defined as the ratio of the temporal widths of the pulse at $x = 10$ µm and $x = 0$, exhibits a rapid increase in the interval $f = 4.5 – 4.8$ GHz, reaches the maximum value of three at $f = 4.8$ GHz, and decreases again at larger frequencies approaching one at $f = 5.2$ GHz [44].

As is known from the theory of waves in dispersive media [30], the strength of dispersion-induced broadening is determined by the rate of change of the group velocity $v_g$ with the variation of the frequency $f$, i.e., $dv_g/df$. We experimentally determine $v_g$ ($f$) by analyzing the spatial dependences of the propagation delay of the spin-wave pulses [Fig. 3(a)] obtained from the Gaussian fits of the temporal profiles [Fig. 2(a)]. As seen from Fig. 3(a), the propagation delay exhibits a well-defined linear spatial dependence enabling an accurate determination of the spin-wave group velocity $v_g$ for different frequencies [Fig. 3(b)] [44]. The obtained data show that the group velocity monotonically decreases with increasing frequency. However, the rate $dv_g/df$ depends on $f$ non-monotonically. In the frequency range $f = 4.5$-$4.6$ GHz, the group velocity changes slowly, which results in the relatively weak dispersion-induced broadening at these frequencies [Fig. 2(c)]. In the range $f = 4.6$-$5.0$ GHz, the group velocity changes most quickly, which results in the enhanced broadening. Finally, at $f > 5.0$ GHz, the dependence $v_g(f)$ begins to plateau and hence the dispersion effects become very weak.

To address the peculiarities of the $v_g(f)$ dependence in more detail, we experimentally determine the dispersion relation for spin waves $f(k)$, where $k$ is the spin-wave wavevector. For this, we perform phase-resolved µ-BLS measurements [24], which allow one to record spatial maps of the spin-wave phase $\varphi$ corresponding to a particular frequency $f$. These measurements are performed by using long microwave pulses with a duration of 160 µs, for which the effects of dispersion can be neglected. Figure 4(a) shows a representative phase map recorded at the frequency $f = 4.8$ GHz. As seen from these data, the phase does not change across the transverse section of the waveguide ($z$-axis) and $\cos(\varphi)$ exhibits well-defined single-period oscillations in the direction along the waveguide axis ($x$-axis). This indicates a single-mode propagation regime of spin waves and allows precise determination of the spin-wave wavelength, which is equal to the spatial period of the oscillations of $\cos(\varphi)$. By repeating the described measurements at different frequencies and performing Fourier analysis of the phase maps, we determine the spin-wave dispersion relation $f(k)$ [symbols in Fig. 4(b)]. According to the theory of Ref. [45] modified for the case of a stripe waveguides [24], the spin-wave dispersion relation for the fundamental waveguide mode can be written as:

$$f(k) = \frac{\gamma}{2\pi}\sqrt{\left[H_0 + 4\pi MF + \frac{2A}{M}k_{tot}^2\right]\left[H_0 + 4\pi M \frac{k^2}{k_{tot}^2}(1-F) + \frac{2A}{M}k_{tot}^2\right]}. \quad (1)$$

Here, $\gamma$ is the gyromagnetic ratio ($\gamma/2\pi = 2.8$ MHz/Oe), $F = (1 - \exp(-k_{tot} d))/( -k_{tot} d)$, $d$ is the film thickness, $k_{tot}^2 = k^2 + (\pi/w)^2$ is the total spin-wave wavevector, $w$ is the waveguide width, and $A$ is the exchange constant.

We emphasize that the experimental data show very good agreement with calculations based on the analytical theory [solid curve in Fig. 4(b)]. This allows us to use the calculated dependence for the analysis in a broad frequency range extending to frequencies, which cannot be addressed experimentally due to the limited efficiency of inductive excitation of spin waves with short wavelengths.

By differentiating the calculated dispersion relation $f(k)$, we obtain the frequency dependence of the group velocity $v_g(f)$ [solid curve in Fig. 4(c)], which shows good agreement with the experimental one [Fig. 3(b)]. The calculations demonstrate that $v_g(f)$ exhibits a non-monotonous behavior resulting in the appearance of a maximum at $f = f^{max} = 4.55$ GHz and a broad minimum at $f = f^{min} = 5.15$ GHz. This non-monotony arises as a result of competition of the dipolar and the exchange interaction, whose contributions to the dispersion of spin waves have different frequency (wavelength) dependence. In particular, the long-range dipolar interaction dominates at small frequencies (large wavelengths), while the exchange interaction is dominant at large spin-wave frequencies (small wavelengths).

This is demonstrated by the dashed and the dash-dot curve in Fig. 4(c), which show the frequency dependences of the group velocity, calculated by taking into account only dipolar and only exchange interaction, respectively. According to Eq. (1), in the absence of the exchange interaction ($A = 0$), the dispersion relation is given by $f(k) = \frac{\gamma}{2\pi}\sqrt{[H_0 + 4\pi MF][H_0 + 4\pi M \frac{k^2}{k_{tot}^2}(1-F)]}$. By differentiating this relation, we obtain the frequency dependence of the spin-wave group velocity shown in Fig. 4(c) by the dashed curve. This dependence exhibits a maximum close to the frequency $f^{max}$, which depends on the ratio between the film thickness and the waveguide width. At frequencies above $f^{max}$, the group velocity quickly decreases reflecting the decrease of the contribution of the dipolar interaction at small wavelengths. In contrast, the contribution to the group velocity of the exchange interaction, which can be described by the simple analytical expression $v_g^{ex} = \frac{4\gamma A}{M}k_{tot}$, is negligible at small frequencies (large wavelengths) and monotonously increases with the increase of the frequency [dash-dot curve in Fig. 4(c)]. As a result of the competition of these effects, a broad minimum of the group velocity is formed at $f = f^{min}$.

Since $dv_g/df \approx 0$ for the frequencies in the vicinity of $f^{max}$ and $f^{min}$, the related dispersion broadening effects are weak for spin-wave pulses with these carrier frequencies [Fig. 2(c)]. Note that the minimum at $f^{min}$ is significantly broader than the maximum at $f^{max}$, and hence short spin-wave pulses excited at frequencies close to $f^{min}$ exhibit weaker, almost negligible dispersion broadening. We emphasize that the group velocity starts to change rapidly again at frequencies $f > f^{min}$ [Fig. 4(c)]. Therefore, one can expect that, at high frequencies, the dispersion-induced broadening comes into play again. Due to the strong reduction of the excitation efficiency, we cannot prove this hypothesis experimentally. To gain information about the propagation of short spin-wave pulses at large frequencies, we perform micromagnetic simulations using the software package MuMax3 [46].

We consider a computation domain with dimensions of $50 \times 1 \times 0.05$ μm³ discretized into $10 \times 10 \times 10$ nm³ cells. The standard value for the YIG exchange constant of $A = 3.66\times10^{-7}$ erg/cm is used [47]. Spin-wave pulses are excited by applying a 2.5 ns long Gaussian pulse of the out-of-plane magnetic field with the amplitude of 0.1 Oe in the center of the computational domain, and the propagation of the pulse is analyzed in the same way, as it is done in the experiment. Figure 5 shows the frequency dependence of the broadening factor, which was defined above, obtained from the simulations. The results of calculations for $f$ = 4.5-5.2 GHz show reasonable agreement with the experimental data of Fig. 2(c). Additionally, at $f >$ 5.2 GHz, we observe an increase of the broadening factor, which confirms our hypothesis that the frequency $f = f^{min}$ corresponds to an optimal propagation regime, where the detrimental dispersion effects are minimized.

Since the optimum frequency is governed by the competition of the dipolar and the exchange interaction, it can be controlled by tuning the geometrical parameters of the waveguide, which influence the dipolar magnetic fields. The most efficient control can be achieved by varying the thickness of the YIG film $d$. This is illustrated in Fig. 6(a) showing frequency dependences of the group velocity calculated for YIG waveguides with different thicknesses. As can be seen from Figs. 6(a) and 6(b) (solid circles), the increase of the thickness results in the shift of $f^{min}$ towards higher frequencies corresponding to smaller wavelengths. This shift is explained by the increase of the contribution of the magnetic dipolar interaction in thicker films: with the increase of the thickness, the maximum of the dipole-dominated group velocity (see dashed curve in Fig. 4(c)) increases, while the exchange contribution (dash-dot curve in Fig. 4(c)) remains unchanged, resulting in the shift of $f^{min}$. In particular, at $d$ = 30 nm, the optimum corresponds to the wavelength of 1 μm, while, at $d$ = 70 nm, it corresponds to the wavelength of 0.45 μm. This indicates that the YIG waveguide can be easily optimized to achieve the optimum propagation for the spin-wave wavelength required for the particular application. We also note that the shift of the optimum is not expected to be accompanied by a significant increase of the dispersion-induced broadening under the optimum conditions. This is illustrated by Fig. 6(b) (open circles) that shows the relative deviation of the group velocity $\Delta v_g = 2(v_g^{max} - v_g^{min})/(v_g^{max} + v_g^{min})$ in a 400 MHz-wide frequency interval approximately corresponding to width of the frequency spectrum of a 2.5 ns-long spin-wave pulse. As seen from this data, the deviation $\Delta v_g$ remains below 10% within the entire interval $d$=30-70 nm.

## IV. CONCLUSIONS

In this study, we have shown that the detrimental dispersion broadening of nanosecond-long spin-wave pulses propagating in ultrathin YIG waveguides can be avoided by using the competitive effects of the dipolar and the exchange interaction on the spin-wave dispersion. This competition results in a formation of the dispersion-free region, where short spin-wave pulses propagate without changing their shape. The region can be efficiently controlled by varying the geometry of the waveguide, which allows one to achieve the dispersionless propagation at different frequencies/wavelengths of spin waves depending on the requirements of the particular application. Our findings should facilitate the implementation of high-speed magnonic circuits capable of handling high data rates.

## ACKNOWLEDGMENTS

This work was supported in part by the Deutsche Forschungsgemeinschaft (DFG, German Research Foundation) – Project-ID 433682494 – SFB 1459, by the French National Research Agency (ANR), ANR MAESTRO project, Grant No. 18-CE24-0021 and as part of the "Investissements d'Avenir" program (Labex NanoSaclay, reference: ANR-10-LABX-0035 "SPiCY" and by the Ile-de-France region SESAME (IMAGESPIN project No. EX039175).

[1] Y. Sun, Y. Y. Song, H. Chang, M. Kabatek, M. Jantz, W. Schneider, M. Wu, H. Schultheiss, and A. Hoffmann, Growth and ferromagnetic resonance properties of nanometer-thick yttrium iron garnet films, Appl. Phys. Lett. **101**, 152405 (2012).

[2] O. d'Allivy Kelly, A. Anane, R. Bernard, J. Ben Youssef, C. Hahn, A. H. Molpeceres, C. Carretero, E. Jacquet, C. Deranlot, P. Bortolotti, R. Lebourgeois, J.-C. Mage, G. de Loubens, O. Klein, V. Cros, and A. Fert, Inverse spin Hall effect in nanometer-thick yttrium iron garnet/Pt system, Appl. Phys. Lett. **103**, 082408 (2013).

[3] C. Hauser, T. Richter, N. Homonnay, C. Eisenschmidt, M. Qaid, H. Deniz, D. Hesse, M. Sawicki, S. G. Ebbinghaus, and G. Schmidt, Yttrium iron garnet thin films with very low damping obtained by recrystallization of amorphous material, Sci. Rep. **6**, 20827 (2016).

[4] S. Neusser and D. Grundler, Magnonics: Spin Waves on the Nanoscale, Adv. Mater. **21**, 2927 (2009).

[5] V. V. Kruglyak, S. O. Demokritov, and D. Grundler, Magnonics, J. Phys. D: Appl. Phys. **43**, 264001 (2010).

[6] B. Lenk, H. Ulrichs, F. Garbs, M. Münzenberg, The building blocks of magnonics, Phys. Rep. **507**, 107–136 (2011).

[7] A. V. Chumak, V. I. Vasyuchka, A. A. Serga, and B. Hillebrands, Magnon spintronics. Nat. Phys. **11**, 453 (2015).

[8] A. Khitun, M. Bao, and K. L. Wang, Magnonic logic circuits, J. Phys. D **43**, 264005 (2010).

[9] S. Dutta, S.-C. Chang, N. Kani, D. E. Nikonov, S. Manipatruni, I. A. Young, and A. Naeemi, Non-volatile Clocked Spin Wave Interconnect for Beyond-CMOS Nanomagnet Pipelines, Sci. Rep. **5**, 9861 (2015).

[10] G. Csaba, A. Papp, and W. Porod, Perspectives of using spin waves for computing and signal processing, Phys. Lett. A **381**, 1471 (2017).

[11] A. Mahmoud, F. Ciubotaru, F. Vanderveken, A. V. Chumak, S. Hamdioui, C. Adelmann, and S. Cotofana, Introduction to spin wave computing, J. Appl. Phys. **128**, 161101 (2020).

[12] S. Li, W. Zhang, J. Ding, J. E. Pearson, V. Novosad, and A. Hoffmann, Epitaxial patterning of nanometer-thick Y3Fe5O12 films with low magnetic damping, Nanoscale **8**, 388 (2016).

[13] Q. Wang, B. Heinz, R. Verba, M. Kewenig, P. Pirro, M. Schneider, T. Meyer, B. Lägel, C. Dubs, T. Brächer, and A. V. Chumak, Spin pinning and spin-wave dispersion in nanoscopic ferromagnetic waveguides, Phys. Rev. Lett. **122**, 247202 (2019).

[14] F. Heyroth, C. Hauser, P. Trempler, P. Geyer, F. Syrowatka, R. Dreyer, S.G. Ebbinghaus, G. Woltersdorf, and G. Schmidt, Monocrystalline freestanding three-dimensional yttrium-iron-garnet magnon nanoresonators, Phys. Rev. Appl. **12**, 054031 (2019).

[15] B. Heinz, T. Brächer, M. Schneider, Q. Wang, B. Lägel, A. M. Friedel, D. Breitbach, S. Steinert, T. Meyer, M. Kewenig, C. Dubs, P. Pirro, and A. V. Chumak, Propagation of spin-wave packets in individual nanosized yttrium iron garnet magnonic conduits. Nano Letters **20**, 4220−4227 (2020).

[16] H. Yu, O. d'Allivy Kelly, V. Cros, R. Bernard, P. Bortolotti, A. Anane, F. Brandl, R. Huber, I. Stasinopoulos, and D. Grundler, Magnetic thin-film insulator with ultra-low spin wave damping for coherent nanomagnonics. Sci. Rep. **4**, 6848 (2014).

[17] M. Collet, O. Gladii, M. Evelt, V. Bessonov, L. Soumah, P. Bortolotti, S. O. Demokritov, Y. Henry, V. Cros, M. Bailleul, V. E. Demidov, and A. Anane, Spin-wave propagation in ultra-thin YIG based waveguides, Appl. Phys. Lett. **110**, 092408 (2017).

[18] C. Liu, J. Chen, T. Liu, F. Heimbach, H. Yu, Y. Xiao, J. Hu, M. Liu, H. Chang, T.Stueckler, S. Tu, Y. Zhang, Y. Zhang, P. Gao, Z. Liao, D. Yu, K. Xia, N. Lei, W. Zhao, and M. Wu, Long-distance propagation of short-wavelength spin waves. Nat. Commun. **9**, 738 (2018).

[19] N. Loayza, M. B. Jungfleisch, A. Hoffmann, M. Bailleul, and V. Vlaminck, Fresnel diffraction of spin waves. Phys. Rev. B **98**, 144430 (2018).

[20] J. Förster, J. Gräfe, J. Bailey, S. Finizio, N. Träger, F. Groß, S. Mayr, H. Stoll, C. Dubs, O. Surzhenko, N. Liebing, G. Woltersdorf, J. Raabe, M. Weigand, G. Schütz, S. Wintz. Direct observation of coherent magnons with suboptical wavelengths in a single-crystalline ferrimagnetic insulator. Physical Review B **100**, 214416 (2019).

[21] B. Divinskiy, N. Thiery, L. Vila, O. Klein, N. Beaulieu, J. Ben Youssef, S. O. Demokritov, and V. E. Demidov, Sub-micrometer near-field focusing of spin waves in ultrathin YIG films, Appl. Phys. Lett. **116**, 062401 (2020).

[22] P. Che, K. Baumgaertl, A. Kúkol'ová, C. Dubs, and D. Grundler, Efficient wavelength conversion of exchange magnons below 100 nm by magnetic coplanar waveguides. Nat. Commun. **11**, 1445 (2020).

[23] I. Bertelli, J. J. Carmiggelt, T. Yu, B. G. Simon, C. C. Pothoven, G. E. W. Bauer, Y. M. Blanter, J. Aarts, and T. van der Sar, Magnetic resonance imaging of spin-wave transport and interference in a magnetic insulator. Sci. Adv. **6**, eabd3556 (2020).

[24] V. E. Demidov and S. O. Demokritov, Magnonic waveguides studied by micro-focus Brillouin light scattering, IEEE Trans. Mag. **51**, 0800215 (2015).

[25] M. Evelt, V. E. Demidov, V. Bessonov, S. O. Demokritov, J. L. Prieto, M. Munoz, J. Ben Youssef, V. V. Naletov, G. de Loubens, O. Klein, M. Collet, K. Garcia-Hernandez, P.

Bortolotti, V. Cros, and A. Anane, High-efficiency control of spin-wave propagation in ultra-thin yttrium iron garnet by the spin-orbit torque. Appl. Phys. Lett. **108**, 172406 (2016).

[26] M. Collet, X. de Milly, O. d'Allivy Kelly, V.V. Naletov, R. Bernard, P. Bortolotti, J. Ben Youssef, V.E. Demidov, S.O. Demokritov, J.L. Prieto, M. Munoz, V. Cros, A. Anane, G. de Loubens, and O. Klein, Generation of coherent spin-wave modes in yttrium iron garnet microdiscs by spin–orbit torque, Nat. Commun. **7**, 10377 (2016).

[27] H. Zhang, M. J. H. Ku, F. Casola, C. H. R. Du, T. van der Sar, M. C. Onbasli, C. A. Ross, Y. Tserkovnyak, A. Yacoby, and R. L. Walsworth, Spin-torque oscillation in a magnetic insulator probed by a single-spin sensor. Phys. Rev. B **102**, 024404 (2020).

[28] M. Evelt, L. Soumah, A. B. Rinkevich, S. O. Demokritov, A. Anane, V. Cros, J. Ben Youssef, G. de Loubens, O. Klein, P. Bortolotti, and V. E. Demidov, Emission of coherent propagating magnons by insulator-based spin-orbit-torque oscillators, Phys. Rev. Appl. **10**, 041002 (2018).

[29] A. G. Gurevich and G. A. Melkov, *Magnetization Oscillations and Waves* (CRC, New York, 1996).

[30] G. P. Agrawal, *Fiber-optic communication systems* (Wiley, 2010).

[31] K. Perzlmaier, G. Woltersdorf, and C. H. Back, Observation of the propagation and interference of spin waves in ferromagnetic thin films, Phys. Rev. B **77**, 054425 (2008).

[32] T. Satoh, Y. Terui, R. Moriya, B. A. Ivanov, K. Ando, E. Saitoh, T. Shimura, and K. Kuroda, Directional control of spin-wave emission by spatially shaped light, Nat. Photonics **6**, 662–666 (2012).

[33] Y. Au, M. Dvornik, T. Davison, E. Ahmad, P. S. Keatley, A. Vansteenkiste, B. Van Waeyenberge, and V. V. Kruglyak, Direct excitation of propagating spin waves by focused ultrashort optical pulses, Phys. Rev. Lett. **110**, 097201 (2013).

[34] S. Iihama, Y. Sasaki, A. Sugihara, A. Kamimaki, Y. Ando, and S. Mizukami, Quantification of a propagating spin-wave packet created by an ultrashort laser pulse in a thin film of a magnetic metal, Phys. Rev. B **94**, 020401(R) (2016).

[35] P. Wessels, A. Vogel, J.-N. Tödt, M. Wieland, G. Meier, and M. Drescher, Direct observation of isolated Damon-Eshbach and backward volume spin-wave packets in ferromagnetic microstripes, Sci. Rep. **6**, 22117 (2016).

[36] S. Muralidhar, R. Khymyn, A. A. Awad, A. Alemán, D. Hanstorp, and J. Åkerman, Femtosecond laser pulse driven caustic spin wave beams, Phys. Rev. Lett. **126**, 037204 (2021).

[37] M. Covington, T. M. Crawford, and G. J. Parker, Time-resolved measurement of propagating spin waves in ferromagnetic thin films, Phys Rev Lett. **89**, 237202 (2002).

[38] Z. Liu, F. Giesen, X. Zhu, R. D. Sydora, and M. R. Freeman, Spin wave dynamics and the determination of intrinsic Gilbert damping in locally-excited Permalloy thin films, Phys. Rev. Lett. **98**, 087201 (2007).

[39] K. Sekiguchi, K. Yamada, S. M. Seo, K. J. Lee, D. Chiba, K. Kobayashi, and T. Ono, Nonreciprocal emission of spin-wave packet in FeNi film, Appl. Phys. Lett. **97**, 022508 (2010).

[40] M. Jamali, J. H. Kwon, S.-M. Seo, K.-J. Lee, and Hyunsoo Yang, Spin wave nonreciprocity for logic device applications, Sci. Rep. **3**, 3160 (2013).

[41] B. Divinskiy, V. E. Demidov, S. O. Demokritov, A. B. Rinkevich, and S. Urazhdin, Route toward high-speed nano-magnonics provided by pure spin currents", Appl. Phys. Lett. **109**, 252401 (2016).

[42] R. Kohno, N. Thiery, K. An, P. Noel, L. Vila, V. V. Naletov, N. Beaulieu, J. Ben Youssef, G. de Loubens, and O. Klein, Enhancement of YIG|Pt spin conductance by local Joule annealing, Appl. Phys. Lett. **118**, 032404 (2021).

[43] V. E. Demidov, M. P. Kostylev, K. Rott, P. Krzysteczko, G. Reiss, and S. O. Demokritov, Excitation of microwaveguide modes by a stripe antenna, Appl. Phys. Lett. **95**, 112509 (2009).

[44] See Supplemental Material at … for the dispersion-free region at elevated static magnetic fields.

[45] B. A. Kalinikos, IEE Proc. H **127**, 4 (1980).

[46] A. Vansteenkiste, J. Leliaert, M. Dvornik, M. Helsen, F. Garcia-Sanchez, and B. Van Waeyenberge, AIP Adv. **4**, 107133 (2014).

[47] S. Klingler, A. V. Chumak, T. Mewes, B. Khodadadi, C. Mewes, C. Dubs, O. Surzhenko, B. Hillebrands, and A. Conca, Measurements of the exchange stiffness of YIG films using broadband ferromagnetic resonance techniques, J. Phys. D: Appl. Phys. **48**, 015001 (2015).

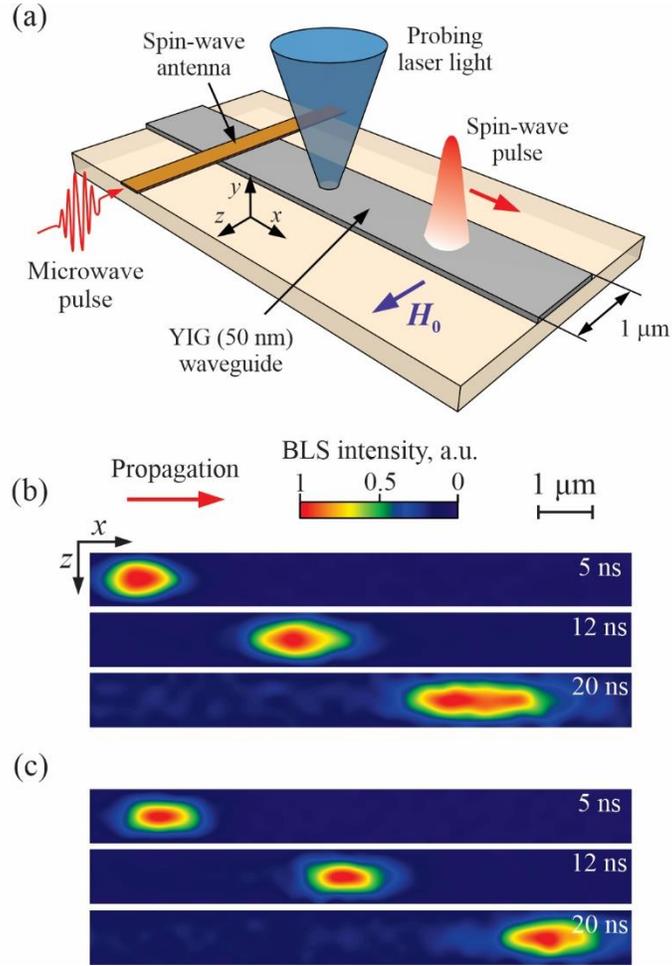

FIG. 1. (a) Schematics of the experiment. (b) and (c) Snapshots of the normalized BLS intensity recorded at different delays relative to the start of the driving microwave pulse, as labelled. (b) $f$=4.8 GHz. (c) $f$=4.5 GHz. The data were obtained at $H_0$ = 1000 Oe.

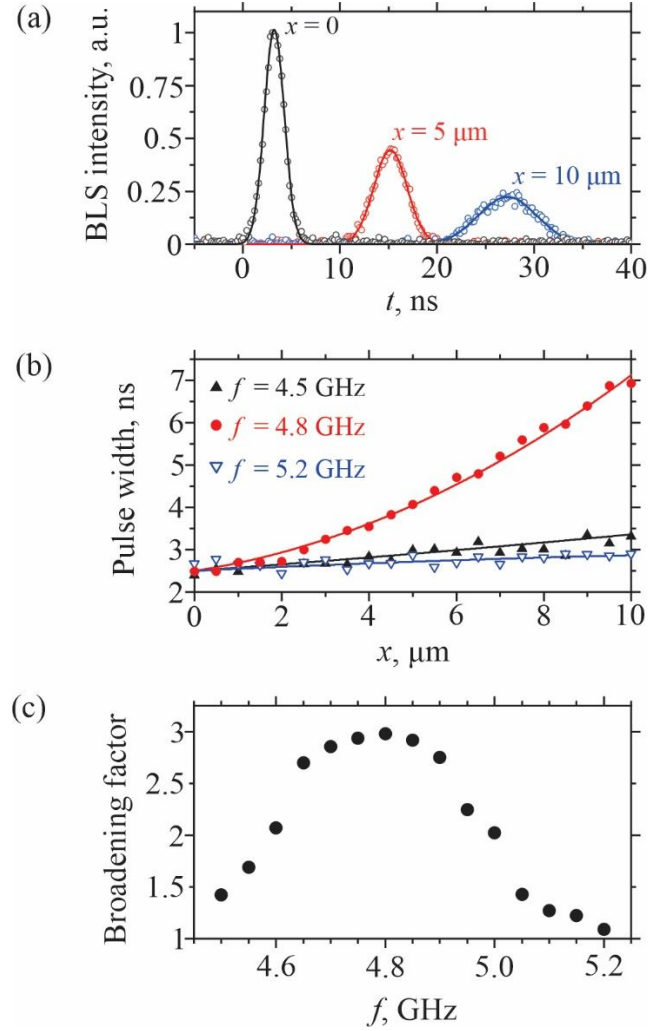

FIG. 2. (a) Temporal profiles of the spin-wave pulse with a carrier frequency $f = 4.8$ GHz recorded at different propagation distances, as labelled. The spin-wave intensity is integrated over the transverse section of the waveguide. Symbols – experimental data. Curves – results of the fitting by a Gaussian function. (b) Spatial dependence of the temporal width of the spin-wave pulse for different carrier frequencies, as labeled. Symbols – experimental data. Curves – guides for the eye. (c) Frequency dependence of the broadening factor, defined as the ratio of the temporal widths of the pulse at $x = 10$ µm and $x = 0$. The data were obtained at $H_0 = 1000$ Oe.

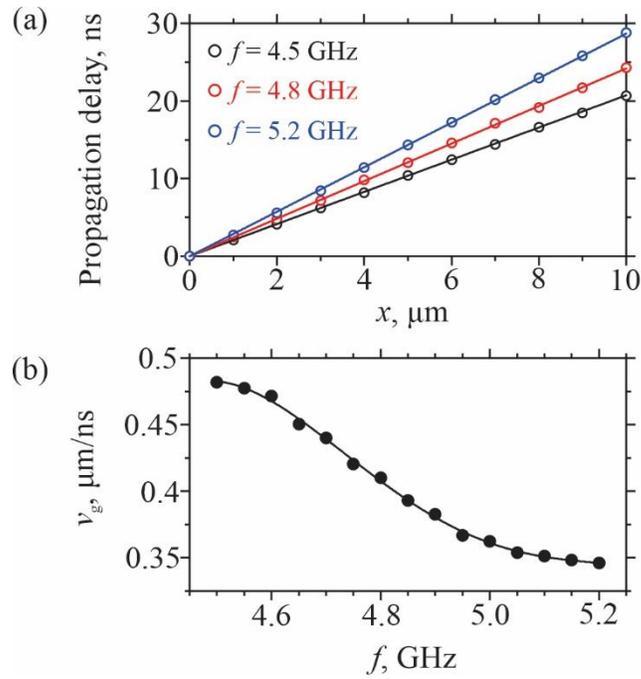

FIG. 3. (a) Spatial dependence of the propagation delay for different carrier frequencies, as labeled. Symbols – experimental data. Lines are linear fits of the experimental data. (b) Group velocity as a function of the carrier frequency. Symbols – experimental data. Curve is the guide for the eye. The data were obtained at $H_0 = 1000$ Oe.

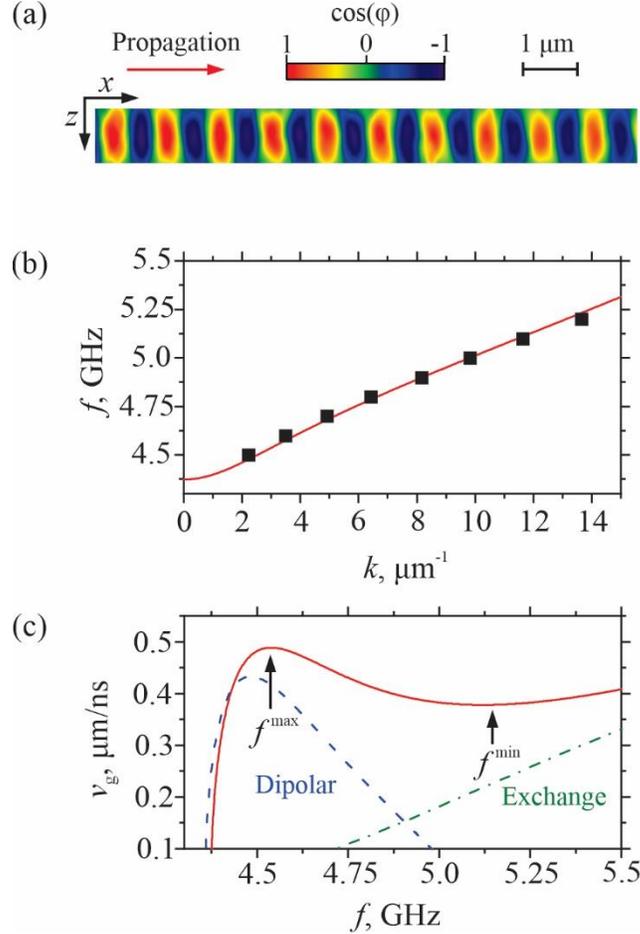

FIG. 4. (a) Representative spatial map of the spin-wave phase recorded at $f = 4.8$ GHz. (b) Spin-wave dispersion relation. Symbols are data extracted from Fourier analysis of phase maps, such as shown in (a). Solid curve is the analytically calculated dispersion relation of spin waves. (c) Solid curve is the calculated frequency dependence of the spin-wave group velocity. $f^{max}$ and $f^{min}$ mark the frequencies of the maximum and the minimum of $v_g(f)$, respectively. Dashed (dash-dotted) curve is the frequency dependence of the group velocity calculated taking into account the dipolar (exchange) interaction only. All data correspond to $H_0 = 1000$ Oe.

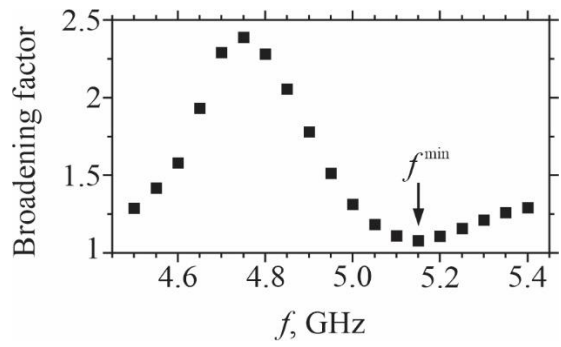

FIG. 5. Frequency dependence of the broadening factor of the spin-wave pulse obtained from micromagnetic simulations. The simulations were performed at $H_0 = 1000$ Oe.

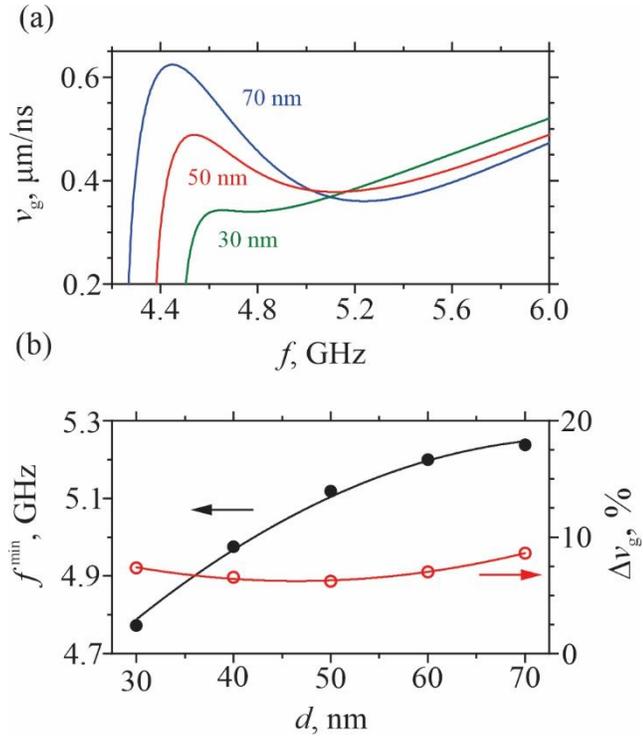

FIG. 6. (a) Frequency dependences of the group velocity calculated for YIG waveguides with different thicknesses, as labelled. (b) Dependences of the frequency $f^{\min}$ and of the relative deviation of the group velocity over a 400 MHz-wide frequency interval on the thickness of the YIG waveguide. All calculations were performed at $H_0 = 1000$ Oe.

## Supplementary information

**Dispersion-free region at elevated static magnetic fields.**

To show that the observed dispersionless propagation can be achieved in a broad range of static magnetic fields, we perform additional measurements at $H_0 = 2000$ Oe. The obtained frequency dependences of the group velocity and of the broadening factor are shown in Fig. 1(a) and 1(b), respectively. The data are qualitatively similar to those obtained at $H_0 = 1000$ Oe (Figs. 3(b) and 2(c) in the main text): one can clearly see the formation of the dispersion-free region at frequencies around 7.95 GHz. We note that, at large static fields, the dispersion-free region slightly shifts toward larger wavelengths, which can be excited by the used spin-wave antenna. This allows the direct experimental observation of the minimum in both dependences shown in Fig. 1.

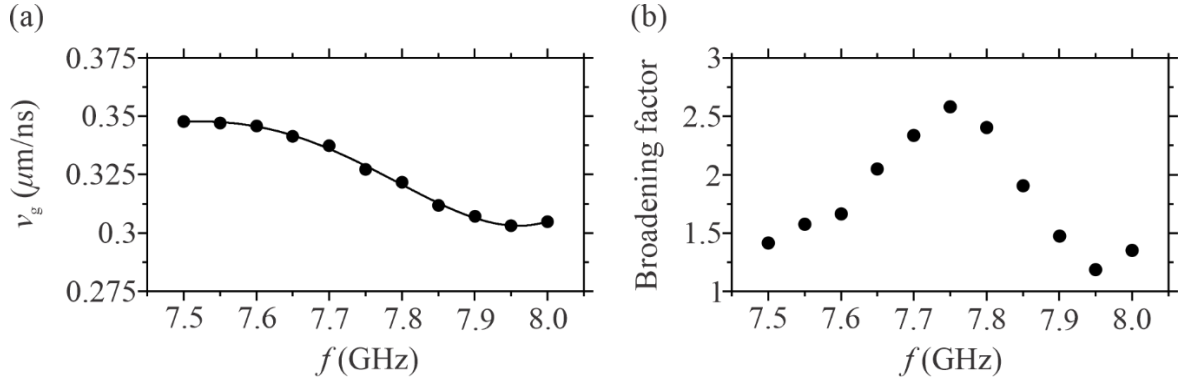

**FIG. 1.** (a) Group velocity as a function of the carrier frequency. Symbols – experimental data. Curve is the guide for the eye. (b) Frequency dependence of the broadening factor, defined as the ratio of the temporal widths of the pulse at $x = 10$ µm and $x = 0$. The data were obtained at $H_0 = 2000$ Oe.